\documentclass[floats,aps,twocolumn,showpacs]{revtex4}
\usepackage{dcolumn}

\usepackage{epsfig}
\usepackage{longtable}
\usepackage{graphicx}
\usepackage{dcolumn}

\begin{document}


\title{Alpha-Cluster Model, Charge Symmetry of Nuclear Force and
Single Particle Bound State Potential in Symmetrical Nuclei
}


\author{G. K. Nie}
\email[]{galani@Uzsci.net}
\affiliation{\it{Institute of Nuclear Physics, Uzbekistan}}


\date{\today}

\begin{abstract}
A phenomenological $\alpha$-cluster model based on $np$-pair interactions and the charge symmetry of nuclear force allows one to estimate the Coulomb energy, the Coulomb radius $R_C$, the Coulomb  energy of the last proton interaction with the residual nucleus and the radius of its position $R_p$ in a symmetrical nucleus. The values $R_C$ and $R_p$ obtained for the symmetrical nuclei with $5\le Z\le 45$ are used in a long standing task of determination of the parameters of the Woods-Saxon potential of neuton/proton bound state used in DWBA analysis of direct one nucleon transfer reactions. According to the charge symmetry of nuclear force a requirement of equality of the nuclear potentials of the last neutron and the last proton in a symmetrical nucleus is added to the standard well-depth procedure in solving the Shr\"{o}dinger equation, which makes $R_C$ the crucial parameter to determine the parameters of the nuclear potential and the value of the last proton $rms$ radius $<r_p^2>^{1/2}$ in the wave representation. Comparative analysis of the radii $<r_p^2>^{1/2}$ and $R_p$, $R_C$, the experimental radii and the nuclear potential radii obtained at the calculations shows that for the nuclei with $Z >$16 it is inappropriate to represent a single particle bound state by the Woods-Saxon potential. Using asymptotic coefficient of wave function allows one to estimate errors in obtaining spectroscopic factor caused by using the standard parameters in DWBA analysis of pure peripheral reactions. It is shown that using these may bring an error up to 48$\%$.
\end{abstract}


\maketitle

\section{INTRODUCTION}

It is well known that nuclear properties are described in the framework of three different representations of nucleon arrangements \cite{1}, the mean field representation, the nucleon liquid drop model and the $\alpha$-cluster model. In analysis of one nucleon transfer reactions $A(a,b)B$, where $A=B+n/p$ and $b=a+n/p$, one has to use the mean filed representation where neutrons and protons move independently in the mean field. The nuclear mean field is described by the Woods-Saxon (W-S) potential due to the single particle model \cite{2} with parameters of radius $r_0$ , diffuseness $a_0$ and depth $V_0$ depicting experimental charge distribution. The standard Distorted Wave Born Approximation (DWBA) approach \cite{3}, with its implementation in a well known programm DWUCK\cite{4} and its modifications, is used in the analysis of differential cross sections of one nucleon transfer reactions. The approach suggests that the nucleon wave function is normalized with $S$ which is the spectroscopic factor (SF). In DWBA analysis $S$ is obtained by dividing the experimental cross sections of the reaction at the main scattering peak by the calculated one. The experimental binding energy of the nucleon (neutron/proton) $\varepsilon_{n/p}$ in a nucleus $A$ is used as the eigenvalue in the $\rm Shr\ddot{o}dinger$ equation for the bound state $A=B+n/p$.

The determination of the single particle bound state potential parameters is a long standing task \cite{5,6,7,8,9,10,11,12,13,14,15}. As the main criteria for selecting proper parameters the experimental nuclear radius and charge distribution are usually used \cite{2,5,6,7,8,9} with strong needs of additional model parameters like sizes of internal shells and the binding energies of the nucleons of the shells, which do not make the obtained values valuable. During a few decades in the DWBA analysis so called standard parameters have been used, these are around the values $r_0=1.25$ fm, $a_0=0.65$ fm, which provide close values of the neutron and proton depths  $V_0\approx 50$ MeV in so called well depth procedure adjusting the depth to the values $\varepsilon_{n/p}$ \cite{7,8,9,10,11}. Some articles, for example \cite{10,11}, are devoted to finding so called global parameters of the W-S potential to describe single particle bound states of the nuclei united on some signs like closeness of the numbers of neutrons and protons to the magic ones. The parameters turned out to be close to the standard ones and they provide description of single particle spectra within an accuracy of 1-2 MeV \cite{11}. In this connection the task of finding the realistic values of the parameters which are supposed to be unique for every nucleus as well as an estimation of the error in obtaining SF by means DWBA with standard parameters is actual.

In \cite{12,13,14} the asymptotic normalization coefficients of the bound state (ANC) $C^2_{n/p}$ was introduced in the analysis and the task of finding the proper values of the parameters was redefined for the task of finding proper value of asymptotic coefficient (AC) $b_{n/p}$ of the bound state wave function. According to the approach the cross section of pure peripheral reaction at the main scattering peak of the angular distribution contains an asymptotic part of the nucleon's wave function which for the last neutron and the last proton are
\begin{equation}
\varphi_n(r)_{r\rightarrow\infty}=b_ni^lkh^l_{lj}(ikr)\label{1}
\end{equation}
\begin{equation}
\varphi_p(r)_{r\rightarrow\infty}=b_pW_{-\eta,lj}(2kr)/r\label{2},
\end{equation}
where $h^l_lj(ikr)$ is the spherical Hankel function of the first kind, $W_{-\eta,lj}(2kr)$ is the Whittaker function, $\eta$ is the Coulomb parameter for the bound state, $k =\sqrt{2\mu\varepsilon_{n/p}}$, $\mu$ is the reduced mass for the bound system $A=B+n/p$. AC $b_{n/p}$ is determined by the geometrical parameters of the potential. Then ANC $C^2_{n/p}$ is related with SF and AC as follows
\begin{equation}
C^2_{n/p}=Sb^2_{n/p}\label{3}.
\end{equation}

This approach allows one to obtain ANCs from analysis of experimental cross sections of pure peripheral reactions like $A(d,t)B$ and $B(^3{\rm He},d)A$ \cite{15,16,17}. The value ANC obtained from the reactions does not depend of the potential parameters used in calculation, because in that case $S$ is in inverse proportion to the $b_{n/p}^2$. In \cite{13} it was proposed to obtain AC and corresponding parameters from analysis of experimental cross sections of direct reactions with using a known value ANC. But later it was shown \cite{15} that not all direct reactions are good for the task. In case of reactions like $(p,d)$ and $(d,n)$ there is a strong dependence of the calculated cross sections on the optical potentials used for input and outgoing channels, which does not allow one to solve the task.

In \cite{18,19,20} a method was developed where the parameters are obtained under the condition of the exact equivalence of the neutron and proton potentials for symmetrical and mirror nuclei (EPN condition) due to the charge symmetry of nuclear force. Together with the well- depth standard procedure the EPN condition makes the parameters dependent on each other, so the Coulomb radius $R_C$ becomes the main critical parameter to determine the value of root mean square (rms) radius calculated for the last proton $<r^2_p>^{1/2}$ and AC. Varying one of the parameters, for example diffuseness $a_0$, at $R_C$ fixed, brings some particular values of $r_0$ and $V_0$ so that $<r^2_p>^{1/2}$ changes within 1\% and $b_{n/p}$ changes within a few per cent. Moreover, the ratio of squared neutron and proton ACs for mirror and symmetrical bound states $b^2_n/b^2_p$ is quite a stable value (with variation within 2\%) at a wide variation of the Coulomb radius. Some explanation why the ratio is to be stable is offered in \cite{21}. The value $b^2_n/b^2_p$ can be used to predict the ratio $C^2_n/C^2_p$ for mirror and symmetrical bound states, because the SFs in the framework of the shell model are equal \cite{22}.

In the present work like in \cite{18,19,20} the EPN condition is used to obtain the potential parameters and AC.  Besides $\varepsilon_{n/p}$, which is calculated as the difference between the binding energies of the nuclei $A$ and $B$, some other macroscopic quantities are used. These are the nuclear Coulomb radius $R_C$ and the last proton position radius (LPPR) $R_p$ in the center of mass system (cms) of a symmetrical nucleus $A=B+n/p$. The values $R_C$ and $R_p$ are estimated from analysis of the nuclear binding energies in the framework of an $\alpha$-cluster model based on $np$-pair interactions with using charge symmetry of nuclear force acting between the nucleons belonging to the $\alpha$- cluster matter \cite{23,24,25,26,27,28}. One of the main aims of developing the model was finding proofs of validity of the EPN condition for single particle bound states in symmetrical and mirror nuclei.

The article consists of 5 sections. The second section explains how the values $R_C$ and $R_p$ are obtained. In the third section the potential parameters obtained by means of the EPN method are given. In section 4 there is a discussion on the results. Conclusions are given in section 5.

\section{LPPR OBTAIND IN ALPHA-CLUSTER MODEL}

For the binding energy  and the Coulomb energy of one $\alpha$-cluster the data for the nucleus $\rm^4He$  are taken, these are (absolute values are given)$\varepsilon_{\alpha}=\varepsilon_{\rm^4He}$ ($\varepsilon_{\rm^4He}=$28.296 MeV \cite{29}) and $\varepsilon^C_\alpha$=0.764. According to the representation of an $\alpha$-cluster liquid drop the nuclear matter is incompressible, so the Coulomb energy of a nucleus equals the energy of the $np$-pairs consisting the $\alpha$-clusters. Then for the nucleus with even $Z$, with the number of $\alpha$-clusters $N_\alpha=Z/2$, and for the nucleus with odd $Z_1=Z+1$, the number of $\alpha$-clusters $N_\alpha+0.5$, the empirical value of the Coulomb energy is supposed to be estimated as follows  \cite{26}
\begin{eqnarray}
E^C=\sum^{N_\alpha}_1{(\Delta {E_{np1}}+\Delta{E_{np}})}\nonumber\\
 E^C_{1}=E^{C}+\Delta {E_{np1}},\label{4}
\end{eqnarray}
where $\Delta {E_{np1}}$ and $\Delta{E_{np}}$ are the differences of one $np$-pair neutron and proton binding energies, odd $np$-pairs have index 1,
\begin{equation}
\Delta E_{np}=\varepsilon_n-\varepsilon_p. \nonumber\\
\Delta E_{np1}=\varepsilon_{n1}-\varepsilon_{p1}, \label{5}
\end{equation}
which is right at one condition that is EPN condition.
From the analysis of the experimental binding energies and the Coulomb energies (4) of the lightest nuclei with $Z,Z_1\le 8$, where there are only short links, some other quantities were obtained \cite{23,24,25,26}. These are the binding energy and the Coulomb energy of a cluster-cluster interaction $\varepsilon_{\alpha \alpha}=$2.425 MeV and $\varepsilon^C_{\alpha \alpha}$=1.925 MeV and the Coulomb energy of the interaction of the single $np$-pair in odd $Z_1$ nucleus with the $\alpha$-cluster of its close vicinity $\varepsilon^C_{np \alpha}$=1.001 MeV. Also a simple formula to describe the binding energies of symmetrical nuclei with 6$\le Z,Z_1\le$29 was found \cite{23} $E=N_\alpha\varepsilon_\alpha+3(N_\alpha-2)\varepsilon_{\alpha \alpha}$ for the nuclei with even $Z$ and $E_1=E+14$ MeV for the nuclei with odd $Z_1$, which means that one added $\alpha$-cluster brings three new links with the closest ones and the $np$-pair in $Z_1$ nucleus has links with the three alpha-clusters of its close vicinity. The formula means that the long range Coulomb energy must be compensated by the surface tension energy, which allows one to find a formula to calculate the latter. Thus a successful formula to calculate binding energy for stable and beta-stable nuclei as well as for the nuclei around the stability valley has been found \cite{26}. In the formula the binding energy of $\alpha$-clusters is calculated separately from the energy of excess neutron pairs ($nn$-pairs). The accuracy of the calculation is comparable with the Weizs\"acker formula \cite{30}, but unlike this well known formula the parameters used in the model are not fitting ones, they had been found from analysis of the binding energies of reduced amount of symmetrical nuclei.

If the Coulomb energy $E^C$ is known, the Coulomb radius $R_C$ and $R_{C1}$ can be found from the formula to calculate the Coulomb energy for a charge sphere

\begin{eqnarray}
E^C=3/5\frac{Z^2e^2}{R_C}\nonumber\\
E^C_1=3/5\frac{Z_1^2e^2}{R_{C1}}.\label{6}
\end{eqnarray}

Another important finding proving the $\alpha$-cluster representation is that the nuclear radius for a stable and a beta-stable nucleus is defined by the number of the $\alpha$-clusters rather than by the total number of the nucleons. The simplest formula for nuclear radius is as follows \cite{25,26}
\begin{equation}
R=R_\alpha N_\alpha^{1/3};\nonumber\\
R_1=R_\alpha (N_\alpha+0.5)^{1/3},\label{7}
\end{equation}
where $R_\alpha=R_{\rm^4He}$ ($R_{\rm^4He}$=1.71 fm \cite{31}) for the nuclei with Z=2 and 5$\le Z,Z_1\le$ 10. For the nuclei $\rm^6Li$ and $\rm^8Be$ the Coulomb repulsion prevents the dense packing. For the nuclei with $Z,Z_1>24$ $R_\alpha=$1.595 fm. To widen the number of nuclei to be described, not restricted with the nuclei of $\beta$-stability, the model was developed to the representation of nucleus as a core (a liquid $\alpha$-cluster drop with dissolved $nn$-pairs in it) and a molecule of a few $\alpha$-clusters on its surface \cite{27,28}. The notion of a nuclear molecule on the surface of a core was developed in \cite{32}. The number of $\alpha$-clusters of the molecule is obtained from analysis of the nuclear binding energy. Thus for the nuclei with $10<Z,Z_1\le24$ the radius $R,R_1$ is defined by the sum of the volumes of the surface molecule, presumably the nucleus $^{20}$Ne and $^{23}$Na (in case of $Z_1$ nuclei one excess neutron is glued to the single $np$-pair), and of the growing core consisting of the $\alpha$-clusters of the radius $R_\alpha=$1.595 fm \cite{27}. The surface tension seems to be responsible for existence of core. It was shown that for the most stable nuclei with $Z,Z_1 \ge 24$ the specific density of the core binding energy $\rho$ is an approximately constant value $\rho \approx 2.5$ $\rm MeV/fm^3$  at the number of the surface molecule $\alpha$-clusters equal to three (three and a half with one excess neutron in case of odd $Z_1$ nuclei). This provides an explanation of the particular number of excess neutrons in stable nuclei \cite{27,28}. In the nuclei with smaller amount of excess neutrons the core is smaller (the molecule is bigger), because the number of $nn$- pairs provides the $\rho$ (less than the saturated value) for a smaller number of $\alpha$-clusters. This theory allows one to calculate the radii for the nuclei with $N\ge Z$ from analysis of the binding energy \cite{27}. It explains the phenomenon of a slight increase of  radii of the isotopes of one $Z,Z_1$ with decreasing $A,A_1$. The obvious success of the model in describing binding energies and radii of the nuclei proves the validity of the EPN condition.

A strait consequence of the EPN condition is an assumption that the Coulomb energy of the last proton interaction with the residual nucleus  $E^C_{p}$ equals the difference of the bindings energy of the neutron and proton of one pair
\begin{equation}
E^C_{p}(r)=\Delta E_{np}, \label{8}
\end{equation}

Then the simplest way to obtain the value of LPPR $R_{p}$ and $R_{p1}$ in the cms of the nuclei $A=2Z$ and $A_1=2Z_1$ is given by the formula
\begin{eqnarray}
\Delta E_{np}=\frac{(Z-1)e^2}{R^{A-1}_p}\nonumber\\
\Delta E_{np1}=\frac{(Z_1-1)e^2}{R^{A_1-1}_{p1}}.\label{9}
\end{eqnarray}
where ${R^{A-1}_p}$ is the radius of the last proton position in the cms of the residual nucleus with the mass $A-1$. When $R^{A-1}_p<R_C$  the following equation is used for the Coulomb potential in the standard DWBA (see the manual to DWUCK program \cite{4})
\begin{eqnarray}
\Delta E_{np}=\frac{(Z-1)e^2}{2R_C}(3-(\frac{R_p^{A-1}}{R_C})^2)\nonumber\\
\Delta E_{np1}=\frac{(Z_1-1)e^2}{2R_{C1}}(3-(\frac{R_{p1}^{A_1-1}}{R_{C1}})^2). \label{10}
\end{eqnarray}
The parameter $r^{DWUCK}_{C}$ and $r^{DWUCK}_{C1}$, used in DWUCK program, relate with the Coulomb radius $R_{C}$ and $R_{C1}$by the following formula \cite{4}
\begin{eqnarray}
R_C=r^{DWUCK}_C(A-1)^{1/3}\nonumber\\
R_{C1}=r^{DWUCK}_{C1}(A_1-1)^{1/3}. \label{11}
\end{eqnarray}
Values $R_{p}$ and $R_{p1}$ are related with $R_{p}^{A-1}$, $R_{p1}^{A_1-1}$ as follows
\begin{eqnarray}
R_p=R_p^{A-1}\frac{(A-1)}{A}\nonumber\\
R_{p1}=R_{p1}^{A_1-1}\frac{(A_1-1)}{A_1} .\label{12}
\end{eqnarray}

The values $R_{p,p1}$ obtained by means (9) and (10) with using (12) are given in Table \ref{tab:table1}. For the nuclei with $Z,Z_1\le$8 the values $R_{p,p1}$
obtained by (10) are considerably less than $R_{exp}$, which is out of reason, and they are not presented in the table. For example $R_p=$2.209 fm for $^{16}$O whereas the experimental radius $R_{^{16}\rm O}$=2.718 fm \cite{31}. For light nuclei the representation of nucleus as a charge sphere (6) is not good, therefore the values $R_{C,C_1}$ do not give reasonable results.
\begin{table}[b]
\caption{\label{tab:table1}
The RLPP $R_{p,p1}$ calculated in the framework of $\alpha$-cluster model representation. $E^C$ is the Coulomb energy (4), $R_C$ is the Coulomb radius (6), $\Delta E_{np}=\varepsilon_n-\varepsilon_p$ \cite{29}. The values $R^{(9)}_{p,p1}$ and $R^{(10)}_{p,p1}$ are obtained by means Eq.s (9) and (10) in correspondence with using (12), the values $R^{(13)}_{p,p1}$ are calculated by (13) together with (15), $R^{(14)}_{p,p1}$ are calculated by (14) with (16), $R^{aver}_{p,p1}$ is calculated by (17).}
\begin{ruledtabular}
\begin{tabular}{ccccccccc}
$Z$   &$E^C$&$R_C$  &$\Delta E_{np}$&$R^{(9)}_{p,p1}$&$R^{aver}_{p,p1}$&
$R^{(10)}_{p,p1}$&$R^{(13)}_{p,p1}$&$R^{(14)}_{p,p1}$\\\hline
 5&  5.304 &4.072&  1.851&2.801&2.529&     &2.529&     \\
 6&  8.067 &3.856&  2.763&2.389&2.491&     &2.491&     \\
 7& 11.070 &3.824&  3.003&2.672&2.745&     &2.745&     \\
 8& 14.607 &3.786&  3.537&2.672&2.764&     &2.764&     \\
 9& 18.149 &3.856&  3.542&3.072&3.092&2.888&3.123&3.266 \\
10& 22.170 &3.897&  4.021&3.062&3.086&2.824&3.172&3.263 \\
11& 26.500 &3.945&  4.330&3.174&3.160&2.983&3.218&3.279 \\
12& 31.338 &3.970&  4.838&3.138&3.119&2.885&3.231&3.242 \\
13& 36.397 &4.012&  5.059&3.284&3.271&3.112&3.323&3.378 \\
14& 41.992 &4.033&  5.595&3.226&3.207&2.985&3.310&3.327 \\
15& 47.717 &4.074&  5.725&3.404&3.402&3.262&3.440&3.504 \\
16& 53.896 &4.104&  6.179&3.386&3.388&3.210&3.471&3.483 \\
17& 60.261 &4.144&  6.365&3.513&3.519&3.390&3.549&3.620 \\
18& 67.010 &4.178&  6.749&3.526&3.540&3.390&3.612&3.621 \\
19& 73.940 &4.218&  6.930&3.642&3.661&3.544&3.678&3.764 \\
20& 81.256 &4.253&  7.316&3.646&3.674&3.532&3.731&3.759 \\
21& 88.533 &4.304&  7.277&3.863&3.919&3.816&3.907&4.035 \\
22& 96.182 &4.348&  7.649&3.864&3.930&3.802&3.958&4.031 \\
23&104.090 &4.391&  7.908&3.919&3.967&3.861&3.961&4.081 \\
24&112.316 &4.431&  8.226&3.942&4.001&3.878&4.034&4.092 \\
25&120.814 &4.470&  8.498&3.985&4.031&3.922&4.027&4.144 \\
26&129.615 &4.506&  8.801&4.012&4.068&3.945&4.101&4.158 \\
27&138.698 &4.541&  9.083&4.046&4.089&3.978&4.087&4.202 \\
28&148.173 &4.572&  9.475&4.030&4.081&3.945&4.113&4.187 \\
29&157.727 &4.607&  9.554&4.147&4.198&4.092&4.190&4.314 \\
30&167.606 &4.639&  9.879&4.157&4.217&4.093&4.243&4.316 \\
31&177.643 &4.674& 10.037&4.235&4.290&4.185&4.279&4.407 \\
32&188.033 &4.705& 10.390&4.229&4.294&4.168&4.314&4.401 \\
33&198.243 &4.746& 10.210&4.445&4.528&4.426&4.498&4.660 \\
34&209.196 &4.774& 10.953&4.275&4.368&4.205&4.357&4.544 \\
35& 219.807&4.815& 10.611&4.548&4.637&4.535&4.604&4.774 \\
36& 230.729&4.853& 10.922&4.550&4.651&4.532&4.647&4.777 \\
37& 241.984&4.888& 11.255&4.544&4.621&4.517&4.596&4.750 \\
38& 253.367&4.924& 11.383&4.619&4.704&4.600&4.715&4.799 \\
39& 265.065&4.958& 11.698&4.618&4.697&4.592&4.671&4.828 \\
40& 276.819&4.994& 11.754&4.718&4.804&4.703&4.816&4.894 \\
41&288.649 &5.032& 11.830&4.810&4.909&4.801&4.871&5.056 \\
42&300.617 &5.070& 11.968&4.874&4.982&4.869&4.978&5.102 \\
43&312.857 &5.106& 12.240&4.884&4.985&4.875&4.947&5.134 \\
44&325.385 &5.141& 12.528&4.886&4.999&4.874&4.985&5.139 \\
45&338.023 &5.176& 12.638&4.958&5.060&4.950&5.022&5.211 \\
\end{tabular}
\end{ruledtabular}
\end{table}

The $\alpha$-cluster model also allows one to estimate the values $R_{p,p1}$ upon a suggestion  that the Coulomb energy of the last proton (or $np$-pair) in an even $Z$ nucleus comes from the Coulomb interaction between two protons (two $np$-pairs) in the last $\alpha$-cluster $\varepsilon^C_\alpha$ plus the Coulomb long range interaction of the last $np$-pair with the residual nucleus of the mass number $A-4$, consisting of $N_\alpha$-1 $\alpha$-clusters. For odd $Z_1$ nucleus the energy of the last proton comes
from the energy of the last $np$-pair interaction with the closest $\alpha$-cluster $\varepsilon^C_{np\alpha}$ plus
the energy of its Coulomb long range interaction with the residual nucleus of the mass number A-6 \cite{26}
\begin{eqnarray}
\Delta E_{np}=\varepsilon^C_\alpha+\frac{(Z-2)e^2}{R^{A-4}_p}\nonumber\\
\Delta E_{np1}=\varepsilon^C_{np\alpha}+\frac{(Z_1-3)e^2}{R^{A_1-6}_{p1}},\label{13}
\end{eqnarray}
where $R^{A-4}_p$ and $R^{A_1-6}_{p1}$ are the distances between the last $np$-pair and the cms
of the residual nucleus with mass number $A-4$ and $A_1-6$.

Another formula comes from the $\alpha$-cluster model representation that the last $\alpha$-cluster has
three links with the nearest $\alpha$-clusters. Then the last $\alpha$-cluster Coulomb energy $\Delta E_\alpha$ is equal
to the sum of its own Coulomb energy  $\varepsilon^C_\alpha$, the energy of the three links
with the nearest clusters 3$\varepsilon^C_{\alpha \alpha}$ and
the energy of the long range interaction with the rest $\alpha$-clusters of the nucleus $2(Z-8)e^2/{R^{A-16}_p}$, where $R^{A-16}_p$ is the distance between the last proton and the cms of the remote $\alpha$-clusters with total mass $A-16$. In case of odd $Z_1$ nucleus the Coulomb energy of the last $np$-pair is the sum of the Coulomb energy of its interaction with the three nearest $\alpha$-clusters
$3\varepsilon^C_{np\alpha}$  and the Coulomb energy of the long range interaction with the rest $\alpha$-clusters of the nucleus $(Z_1-7)e^2/{R^{A_1-14}_{p1}}$, where $R^{A_1-14}_p$ is the distance between the last proton and the cms of the remote $\alpha$-clusters with the total mass number $A_1-14$ \cite{26}
\begin{eqnarray}
\Delta E_\alpha=\varepsilon^C_\alpha+3\varepsilon^C_{\alpha\alpha}+\frac{2(Z-8)e^2}{R^{A-16}_p}\nonumber\\
\Delta E_{np1}=3\varepsilon^C_{np\alpha}+\frac{(Z_1-7)e^2}{R^{A_1-14}_{p1}},\label{14}
\end{eqnarray}
where $\Delta E_\alpha=\Delta E_{np1}+\Delta E_{np}$. The simplest formula for the Coulomb energy decreasing with distance is used, as the values  $R^{A-4}_p$, $R^{A_1-6}_{p1}$, $R^{A-16}_{p}$ and $R^{A_1-14}_{p1}$ are surely bigger than the nuclear radius.

LPPRs $R_{p,p1}$ in cms of nucleus $A$ are calculated as follows

\begin{eqnarray}
R_p=R^{A-4}_p\frac{A-4}{A}+R_{^4\rm He}\frac{4}{A}\nonumber\\
R_{p1}=R^{A_1-6}_{p1}\frac{A_1-6}{A_1}+R_{^4\rm He+np}\frac{6}{A_1},\label{15}
\end{eqnarray}
where $R_{^4\rm He}$=1.71 fm, $R_{^4\rm He+np}$=2.57 fm, the experimental radius of $\rm ^7Li$.
To calculate $R_{p,p1}$ in case of (14) one can use LPPR in the nuclei $^{16}$O $R^{16}_p$ and $^{14}$N $R^{14}_{p1}$ obtained by (9), see Table~\ref{tab:table1}. Then using the same logic as in (15) we have
\begin{eqnarray}
R_p=R^{A-16}_p\frac{A-16}{A}+R^{16}_{p}\frac{16}{A}\nonumber\\ R_{p1}=R^{A_1-14}_{p1}\frac{A_1-14}{A_1}+R^{14}_{p1}\frac{14}{A_1}.\label{16}
\end{eqnarray}
The average value $R^{aver}_{p,p1}$ of the radii estimated with using the $\alpha$-cluster model parameters $r_C$, $\varepsilon^C_\alpha$, $\varepsilon^C_{\alpha\alpha}$, $\varepsilon^C_{np\alpha}$ is also calculated as follows
\begin{equation}
R^{aver}_{p,p1}=(R^{(10)}_{p,p1}+R^{(13)}_{p,p1}+R^{(14)}_{p,p1})/3.\label{17}
\end{equation}
The values $R_{p,p1}$ in cms of nuclei $A,A_1$ estimated by different ways are given with upper indexes corresponding to the Eqs, see Table I. All the methods provide radii in a consistent way. Odd $np$-pairs make a leap and the next even $np$-pairs fix the distance. The smallest values of the radii are provided by (10) and the largest ones are given by (14). In calculation of $R^{aver}_{p,p1}$ the values of (10) and (14) approximately compensate each other. So $R^{aver}_{p,p1}$ is in a close agreement with $R_p^{13}$. The difference $|R^{(9)}_{p,p1}-R^{aver}_{p,p1}|$ is little for light nuclei (about 0.02 fm) and increases with $Z,Z_1$ up to 0.1 fm. $R^{(9)}_{p,p1}$ obtained without model parameters is in agreement with $R^{aver}_{p,p1}$ with the rms deviation 0.06 fm.

Both $R^{(9)}_{p,p1}$ and $R^{aver}_{p,p1}$ are consistent with the experimental radii. Not for all symmetrical nuclei there are experimental data. In those cases the experimental radius of the nearest isotope is used \cite{33}. In the case of no data, $Z_1$=43, Eq. (7) is used. The squared nuclear radius ($(R^{A,Z})^2$, $(R^{A_1,Z_1})^2$) is calculated as the sum of the squared radius of the residual nucleus ($(R^{A-1,Z-1})^2$, $(R^{A_1-1,Z_1-1})^2$) and the square radius of the last proton position in the cms of the residual nucleus ($(R^{A-1}_{p})^2$, $(R^{A_1-1}_{p1})^2$) weighed
\begin{eqnarray}
(R^{A,Z})^2=\frac{(Z-1)}{Z}(R^{A-1,Z-1})^2+\frac{1}{Z}(R^{A-1}_p)^2\nonumber\\
(R^{A_1,Z_1})^2=\frac{(Z_1-1)}{Z_1}(R^{A_1-1,Z_1-1})^2+\frac{1}{Z_1}(R^{A_1-1}_{p1})^2.\label{18}
\end{eqnarray}
With increasing $Z,Z_1$ the difference $|R^{(9)}_{p,p1}-R^{aver}_{p,p1}|$ does not affect much the calculated nuclear radii (18). It should be noticed here that for the nuclei with $Z,Z_1=$5, 6, 8, 10, 11 $R^{aver}_{p,p1}$ gives better agreement with the experimental data. For the nuclei with $Z,Z_1\le 8$ the values $R^{aver}_{p,p1}=R^{(13)}_{p,p1}$ (see Table I). In case of the light nuclei the values $R^{aver}_{p,p1}$  are more preferable than $R^{(9)}_{p,p1}$, but the obvious advantage of the latter is that $R^{(9)}_{p,p1}$ is obtained without any parameters. For further analysis the values LPPR $R^{(9)}_{p,p1}$ are used and they are indicated as $R_{p,p1}$ except for the nuclei with $Z,Z_1$=5, 6, then the values $R^{aver}_{p,p1}$ are taken.

\section{BOUND STATE POTENTIAL PARAMETERS IN SYMMETRICAL NUCLEUS}

Parameter $r^{DWUCK}_{C,C1}$ (11) is used together with the parameters of the nuclear W-S potential $r_0$, defined as follows
\begin{equation}
R_{0}=r_0(A-1)^{1/3},\label{19}\nonumber\\
\end{equation}
where $R_0$ is the half-potential radius of the well, and the depth $V_0$ in solving the Shr\"odinger equation for one nucleon bound state with the spin-orbit part of Thomas form with $\lambda$=25 as it is used in DWUCK program \cite{4}
\begin{equation}
V(r)=V_0(f(r)+\frac{\lambda}{45.2}\frac{1}{r}\frac{df(r)}{dr}\vec{L}\vec{\sigma})\label{20}
\end{equation}
where
\begin{equation}
f(r)=[1+exp(\frac{r-r_0(A-1)^{1/3}}{a_0})]^{-1}\label{21}.
\end{equation}
The programm DWUCK was modified with implementation of the EPN condition \cite{18,19}.
The EPN-condition used together with well-depth procedure makes the parameter $r^{DWUCK}_{C,C1}$ the critical one in determination of $r_0$ and $V_0$. At the $r^{DWUCK}_{C,C1}$ fixed, variation of one of the parameters, for example the diffuseness $a_0$ within $0.4\div 0.7$ fm, changes the other parameters $r_0$ and $V_0$ that way that the rms radius $<r_{n/p}^2>^{1/2}$ stays almost the same (variety within 1\%) and AC $b_{n/p}$ changes within a few per cent \cite{19}. So one can say that in such calculations the Coulomb radius defines last proton rms radius. Quantum numbers $n,l,j$ are selected according to the Pauli conservation principle, parity conservation rule and the sum rule for momenta. For those cases when both binding energies and spins of nuclei $A$ and $B$ are known \cite{29}, parameters $r_0$ and $V_0$ of W-S potential have been found at $a_0=0.65$ fm. They are given in Table II.
\begin{table*}[]
\caption{\label{tab:table2}Parameters of the W-S potential at the EPN-condition.
In the first column the bound state $A=B+n/p$; in the 2nd and the 3d columns there are the quantum numbers
$n,l,j$ and experimental binding energies $\varepsilon_{n,p}$\cite{29}; the 4th column gives Coulomb radii $r^{DWUCK}_{C,C1}$ (11); the 5th and the 6th contains the parameters $r_0$ and $V_0$ at $a_0=$0.65 fm and rms radii $<r^2_{n,p}>^{1/2}$; the 7th column contains $R_{p,p1}$ (for Z=5,6 see $R_{p,p1}^{13}$, for the other nuclei $R_{p,p1}^{9}$ in Table \ref{tab:table1}); the 8th column presents AC $b_{n,p}$; the 9th contains AC $b^{st}_{n,p}$ calculated at the standard parameters at the standard well-depth procedure, the 10th column gives the error $\delta$ brought by the standard parameters in DWBA analysis to obtain SF (22).}
\begin{ruledtabular}
\begin{tabular}{cccccccccc}
$A$&$n,l,j$&$\varepsilon_{n,p}$&$r^{DWBA}_{C,C1}$&$r_0$ fm&$<r^2_{n,p}>^{1/2}$&
$R_{p,p1}$&$b_{n,p}$&$b^{st}_{n,p}$&$\delta$\\
$B+n/p$&         &MeV& fm&$V_0$,MeV & fm &
$ fm$&$fm^{-1/2}$&$fm^{-1/2}$&\%\\\hline
$\rm^{10}B=^9$B+$n$&1,1,3/2      & 8.438&      &0.8728  &2.53 &      &2.34  &2.96 & 60 \\
$\rm^{10}B=^9$Be+$p$&            & 6.587&1.958&-80.3740 &2.57 &2.53  &2.60  &3.31 & 62 \\ \hline
$\rm^{12}C=^{11}$C+$n$&1,1,3/2   &18.720&      &1.1520  &2.46 &      &7.81  &8.80 &27  \\
$\rm^{12}C=^{11}$B+$p$&	         &15.957&*1.405 &-69.0963&2.49 &2.49  &8.93  &10.1&28  \\\hline
$\rm^{14}N=^{13}$N+$n$&1,1,1/2   &10.554&      &1.0568  &2.61 &      &3.53  &4.23 &44  \\
$\rm^{14}N=^{13}$C+$p$&          & 7.550&*1.512 &-70.4114&2.67 &2.67  &4.17  &5.02&45  \\	 \hline
$\rm^{16}O=^{15}$O+$n$&1,1,1/2   &15.663&      &1.2032  &2.62 &      &7.14  &7.58 &13  \\
$\rm^{16}O=^{15}$N+$p$&		     &12.128&*1.410 &-61.5594 &2.67 &2.67  &8.88  &9.47&14  \\ \hline                               $\rm^{18}F=^{17}$F+$n$&1,2,5/2   & 9.149&      &1.1370   &3.09 &      &2.37  &2.76 &36  \\                                 $\rm^{18}F=^{17}$O+$p$&		     & 5.607&1.500&-67.8928 &3.15 &3.07  &2.89  &3.38 &37  \\  \hline
$\rm^{20}Ne=^{19}$Ne+$n$&1,0,1/2 &16.865&     &1.1056   &2.99 &      &17.2  &20.9 &48  \\
$\rm^{20}Ne=^{19}$F+$p$ &        &12.844&1.461&-89.0360 &3.04 &3.06  &24.6  &29.8 &47  \\  \hline                          $\rm^{22}Na=^{21}$Na+$n$&1,2,5/2 &11.069&     &1.1541   &3.15 &      &3.69  &4.32 &37  \\                                 $\rm^{22}Na=^{21}$Ne+$p$&        & 6.740&1.430&-63.1768 &3.21 &3.17  &5.12  &6.01 &38  \\ \hline                          $\rm^{24}Mg=^{23}$Mg+$n$&1,2,3/2 &16.531&     &1.2446   &3.02 &      &7.77  &7.85 & 2    \\                                 $\rm^{24}Mg=^{23}$Na+$p$&        &11.693&1.396&-74.9084 &3.06 &3.14  &11.4  &11.6 & 4    \\ \hline                          $\rm^{26}Al=^{25}$Al+$n$&1,2,5/2 &11.365&     &1.1605   &3.26 &      &4.43  &5.15 &35    \\                                          $\rm^{26}Al=^{25}$Mg+$p$&        & 6.306&1.372&-57.8340 &3.33 &3.28  &6.95  &8.10 &36    \\ \hline                          $\rm^{28}Si=^{27}$Si+$n$&1,2,5/2 &17.180&     &1.1390   &3.10 &      &9.36  &11.7 &56    \\                                  $\rm^{28}Si=^{27}$Al+$p$&        &11.585&1.344&-65.949 &3.15 &3.23   &15.3  &19.2 &57    \\  \hline
$\rm^{30}P =^{29}$P +$n$&1,2,3/2 &11.319&     &1.2807   &3.39 &      &5.20  &4.92 &10    \\
$\rm^{30}P =^{29}$Si+$p$&        &5.595 &1.326&-54.8503 &3.49 &3.40  &9.51  &9.04 &10    \\ \hline
$\rm^{32}S =^{31}$S +$n$&2,0,1/2 &15.042&     &1.1866   &3.34 &      &21.9  &24.3 &23    \\
$\rm^{32}S =^{31}$P +$p$&        & 8.863&1.306&-62.2411 &3.45 &3.39  &45.9  &50.8 &22    \\\hline
$\rm^{34}Cl=^{33}$S +$n$&1,2,3/2 &11.508&     &1.2939   &3.51 &      &6.12  &5.63 &15    \\
$\rm^{34}Cl=^{33}$S +$p$&        & 5.143&1.292&-50.3894 &3.63 &3.51  &13.4  &12.4 &14    \\ \hline                          $\rm^{36}Ar=^{35}$Ar+$n$&1,2,3/2 &15.256&     &1.3268   &3.49 &      &11.8  &9.92 &29    \\                                 $\rm^{36}Ar=^{35}$Cl+$p$&        & 8.507&1.277&-52.2729 &3.59 &3.53  &25.9  &21.9 &29    \\\hline
$\rm^{38}K =^{37}$K +$n$&1,2,3/2 &12.072&     &1.3326   &3.66 &      &8.02  &6.77 &29    \\
$\rm^{38}K =^{37}$Ar+$p$&        & 5.143&1.266& -45.9903&3.79 &3.64  &21.2  &18.0 &28    \\ \hline                          $\rm^{40}Ca=^{39}Ca$+$n$&1,2,3/2 &15.644&     &1.3545   &3.63&       &14.8  &11.6 &39    \\
$\rm^{40}Ca=^{39}K $+$p$&        & 8.328&1.254&-48.6024 &3.74 &3.65  &38.1  &29.9 &38    \\\hline
$\rm^{42}Sc=^{41}Sc$+$n$&1,3,7/2 &11.550&     &1.2647   &3.94&       &4.80  &4.64 & 7    \\                                 $\rm^{42}Sc=^{41}Ca$+$p$&        & 4.273&1.248&-56.0627 &4.05 &3.86  &13.9  &13.4 & 7    \\\hline
$\rm^{44}Ti=^{43}Ti$+$n$&1,3,7/2 &16.299&     &1.2881   &3.91&       &12.1  &10.9 &19    \\                                 $\rm^{44}Ti=^{43}Sc$+$p$&        & 8.650&1.241&-59.8610 &4.00&3.86   &32.9  &29.6 &19    \\\hline
$\rm^{46}V =^{45}V $+$n$&1,3,7/2 &13.265&     &1.2599   &3.97&       &7.20  &7.02 & 5    \\
$\rm^{46}V =^{45}Ti$+$p$&        & 5.357&1.235&-56.2452 &4.08 &3.92  &23.8  &23.22& 5   \\\hline
$\rm^{48}Cr=^{47}Cr$+$n$&2,1,3/2 &16.332&     &1.3800   &4.06&       &      &     & 0   \\
$\rm^{48}Cr=^{47}V $+$p$&        & 8.106&1.228&-58.7101 &4.21 &3.94  &      &     & 0   \\\hline
$\rm^{50}Mn=^{49}Mn$+$n$&1,3,5/2 &13.083&     &1.3443   &4.02&       &7.54  &9.75 &67   \\                                 $\rm^{50}Mn=^{49}Cr$+$p$&        & 4.585&1.221&-56.4339 &4.14&3.99   &32.7  &35.8 &20   \\\hline
$\rm^{52}Fe=^{52}Fe$+$n$&1,3,5/2 &16.180&     &1.3669   &4.03&       &14.0  &10.0 &49   \\                                 $\rm^{52}Fe=^{51}Mn$+$p$&        & 7.379&1.215&-57.9168 &4.15&4.01   &55.2  &39.6 &49   \\\hline
$\rm^{54}Co=^{53}Co$+$n$&1,3,7/2 &13.436&     &1.2357   &4.06&       &8.41  &8.74 & 8   \\                                 $\rm^{54}Co=^{53}Fe$+$p$&        & 4.353&1.209&-53.7392 &4.19&4.06   &47.0  &57.6 &50   \\\hline
$\rm^{56}Ni=^{55}Ni$+$n$&1,3,7/2 &16.639&     &1.2359   &4.02&       &14.9  &15.5 & 8   \\                                 $\rm^{56}Ni=^{55}Co$+$p$&        & 7.164&1.202&-57.0697 &4.13&4.03   &70.0  &72.7 & 8   \\\hline
$\rm^{58}Cu=^{57}Cu$+$n$&2,1,3/2 &12.423&     &1.3114   &4.25&       &      &     &    \\                                 $\rm^{58}Cu=^{57}Ni$+$p$&        & 2.869&1.197&-52.1368 &4.52&4.15   &      &     &    \\\hline
$\rm^{60}Zn=^{59}Zn$+$n$&2,1,3/2 &14.998&     &1.3618   &4.26&       &      &     &    \\                                 $\rm^{60}Zn=^{59}Cu$+$p$&        & 5.119&1.192&-52.3728  &4.47&4.16  &      &     &    \\\hline
$\rm^{62}Ga=^{61}Ga$+$n$&2,1,3/2 &12.981&     &1.5085   &4.64&      &       &     &    \\                                 $\rm^{62}Ga=^{61}Zn$+$p$&        & 2.944&1.187&-42.9160 &4.93&4.24  &       &     &    \\\hline
$\rm^{64}Ge=^5{63}Ge$+$n$&2,1,3/2&15.482&     &1.3848   &4.34&      &       &     &    \\                                 $\rm^{64}Ge=^{63}Ga$+$p$&        & 5.092&1.183&-50.5167 &4.55&4.23  &       &     &    \\  \hline                         \end{tabular}
\end{ruledtabular}
\end{table*}
\begin{table*}[]
\begin{ruledtabular}
\begin{tabular}{cccccccccc}
$A$&$n,l,j$&$\varepsilon_{n,p}$&$r^{DWBA}_{C,C1}$&$r_0$ fm&$<r^2_{n,p}>^{1/2}$&
$R_{p,p1}$&$b_{n,p}$&$b^{st}_{n,p}$&$\delta$\\
$B+n/p$&         &MeV& fm&$V_0$,MeV & fm &
$ fm$&$fm^{-1/2}$&$fm^{-1/2}$&\%\\\hline
$\rm^{68}Se=^{67}Se$+$n$&1,3,5/2&15.809&     &1.3346  &4.27&      &18.0&13.7&42\\
$\rm^{68}Se=^{67}As$+$p$&       &4.856 &1.176&-51.6985&4.43 &4.28&&&       \\  \hline
$\rm^{72}Kr=^{73}Kr$+$n$&1,3,5/2&15.089&     &1.4889  &4.77&      &28.6&13.18&79\\
$\rm^{72}Kr=^{73}Br$+$p$&       &4.167 &1.172&-42.2215&4.99&4.55&&&   \\      \hline
$\rm^{74}Rb=^{73}Rb$+$n$&1,3,5/2&13.910&     &1.3916  &4.58&     &17.6&11.19&54  \\
$\rm^{74}Rb=^{73}Kr$+$p$&       &2.655 &1.170&-44.2553&4.79&4.54&&&  \\            \hline
$\rm^{76}Sr=^{75}Y $+$n$&2,1,3/2&15.694&     &1.6178  &4.95&     &&&\\
$\rm^{76}Sr=^{75}Rb$+$p$&       &4.311 &1.168&-39.5242&5.20&4.62&   &  &      \\  \hline
$\rm^{78}Y =^{77}Y $+$n$&2,2,5/2&13.744&     &1.3778  &4.78&      &&&\\
$\rm^{78}Y =^{77}Sr$+$p$&       &2.046 &1.165&-56.2448&5.05&4.62&&&\\\hline
\end{tabular}
\end{ruledtabular}
\end{table*}
In case of the nuclei $\rm^{12}C$ and $\rm^{14}N$ there is not any solution for the $R_C^{DWUCK}=R_C$. It is understandable, because Eq.(6) is not good approach for the light nuclei. In these cases the values $r_C^{DWUCK}$ (marked by *)are fitted to satisfy the demand $<r_p^2>^{1/2}=R_p$. The same was done for $\rm^{16}O$, because the obtained value $<r_p^{2}>^{1/2}=2.25$ fm is less than the radius of the nucleus.

The value $a_0$=0.55 fm in the case of nucleus $^{24}$Mg gives $r_0$=1.2889 fm $V_0$=-68.9871 MeV $<r^2_n>^{1/2}$=3.02fm, $<r^2_p>^{1/2}$=3.06 fm, $b_n$=7.31$fm^{-1/2}$ and $b_p$=10.70$fm^{-1/2}$. Comparison of the values with the data in Table II shows that $a_0$=0.55 fm almost does not change rms radius and ACs change by 6\%. Searching parameters under the EPN condition is done with using iteration. Result insignificantly depends on the starting values in the iteration, which shows the accuracy of the calculation. This is few 0.001 fm for $r_0$ and a few 0.01 MeV for $V_0$, which consequences the rms radius accuracy of 0.001 fm and the AC accuracy of few 0.01 $\rm fm^{-1/2}$.

The procedure of seeking parameters under the EPN-condition brings the determination of rms radius $<r_{p,p1}^2>^{1/2}$  and radius $r_0$ by the Coulomb radius $r^{DWUCK}_{C,C1}$ in a way that decreasing the latter causes increasing $<r_{p,p1}^2>^{1/2}$ and $r_0$.

In the table also the values AC $b^{st}_{n,p}$ calculated at the standard parameters $r_0$=1.25 fm, $a_0$=0.65 fm with $r^{DWUCK}_{C,C1}$=1.25 fm at the standard well-depth procedure without the EPN condition are given. One wants to know what error in obtaining SF in analysis of one nucleon transfer reactions by means of DWBA is brought with using standard parameters. In case of the pure peripheral reactions the cross sections at main scattering peak contains an asymptotic part of the wave function (1) and (2). Then the deviation of the SF S extracted by the DWBA analysis with standard parameters from the value obtained with using the EPN-parameters is calculated as follows
\begin{equation}
\delta=\frac{|b_{n/p}^2-(b^{st}_{n/p})^2|}{b_{n/p}^2}*100\%\label{22}
\end{equation}
It brings some certain error in the extracting spectroscopic information from analysis of direct one nucleon transfer reactions. The total error also includes uncertainty connected with choosing optical potentials to describe incoming and outgoing channels of the reaction, which for the pure peripheral reactions is restricted within 20$\%$.

In the calculations the value $b^2_{n}/b^2_{p}$ obtained under the EPN condition is almost stable value to a wide variety of $r^{DWBA}_{C,C1}$. Varying  $r^{DWBA}_{C,C1}$=1.1-1.5 fm brings a change of the ratio within 2 \%, which can be used to predict the ratio of experimentally obtained values ANC's $C^2_{n}/C^2_{p}$ in case of mirror and symmetrical nuclei due to the equality of the spectroscopic factors of the bound sates\cite{22}.

The demand $<r_p^2>^{1/2}=R_{p,p1}$ can also be taken as a criterion in finding parameters of W-S potential, as it was done in case of nuclei $\rm ^{12}C$, $\rm^{14}N$ and $\rm^{16}O$. Then the Coulomb radius becomes a fitting parameter to find the solution of the Shr\"{o}dinger equation with EPN condition. The calculations have been made for the other nuclei also. For some nuclei the deviation $\delta$ is smaller than in case the criterion is $R_C$. These are the nuclei with $Z,Z_1=10(41\%)$; $14(32\%)$; $15(7\%)$; $17(7\%)$; $18(17\%)$; $19(6\%)$; $20(25\%)$; $22(13\%)$; $26(33\%)$. For example for the nucleus $\rm^{34}Cl$ the following parameters have been obtained: $r^{DWUCK}_{C,C1}$=1.336 fm, $r_0$=1.2394 fm, $V_0$=-54.0661 MeV, $b_p=12.0 fm^{-1/2}$.

\section{DISCUSSION}
In this section for further convenience even $Z$ and odd $Z_1$ are indicated as $Z$, as well as the other values like $A$, $R_p$, $R_C$,$R^{DWUCK}_C$ and $R_0$.

\begin{figure}
\psfig{figure=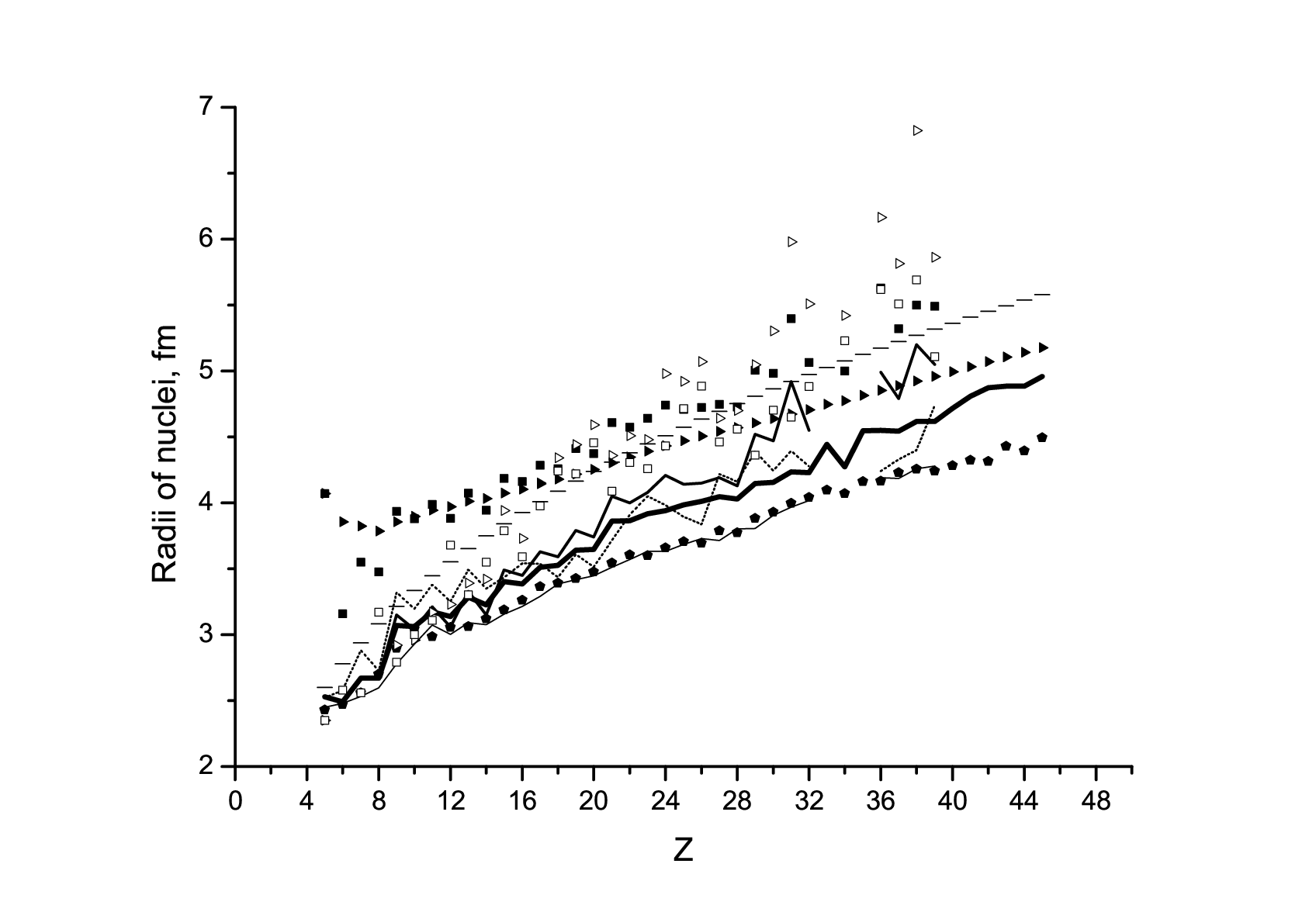,width=0.46\textwidth}
\caption{\label{fig1}Nuclear radii. Symbols indicate the following values: the experimental nuclear radii $R_{exp}$(black pentagons) \cite{33}; the Coulomb radius $R^{DWUCK}_C=R_{C}$ (6) (black triangles); the corresponding radii of W-S nuclear potential $R_0$ satisfying the EPN-condition  (empty triangles); the radii $R^{DWUCK}_C$ obtained under the demand $<r^2>^{1/2}=R_p$ (black squares) and the corresponding radii of nuclear potential $R_0$ (empty squares). Lines indicate: LPPR $R_p$ (solid thick line); rms radii $<r^{2}_p>^{1/2}$ (thin line) calculated with using $R^{DWUCK}_C=R_{C}$ (6) and calculated nuclear radii $R^{A,Z}$ (18)(solid thin line going along the symbols of $R_{exp}$, in Eq. (18) instead $R^{A-1}_p$ the values rms radii $<r^2_p>^{1/2}$ transferred into the c.m.s. of the nucleus A-1 by (12) are used). Also the standard potential radii $R_0=R^{DWUCK}_C=1.25(A-1)^{1/3}$ (dashed line) and the corresponding rms radii $<r^{2}_p>^{1/2}$ obtained in the standard well- depth procedure (line of dots) are given.}
\end{figure}

In FIG.~\ref{fig1} the values of $R_C$ and $R_p$ obtained from alpha-cluster model and the radius of W-S potential $R_0$ satisfying the EPN-condition and the calculated rms radii $<r^2>^{1/2}$ together with the experimental nuclear radii \cite{33} are given. In case of absent data (this is Z=43, A=86), the radius is calculated by (7). There are also the the values $R^{DWUCK}_{C}$ fitted to satisfy equality $<r^2>^{1/2}=R_p$ and the corresponding values $R_{0}$.

The Coulomb radii $R^{DWUCK}_C=R_{C}$ (6), LPPR $R_p$ and experimental nuclear radii $R_{exp}$ \cite{33} are some characteristics of the symmetrical nuclei obtained from the analysis beyond the wave function representation. One can see in FIG.~\ref{fig1} that the Coulomb radius $R_C$ obtained in the framework of the $\alpha$-cluster model relates with $R_{exp}$ as $R_C= R_{exp}+d$ where $d\approx 1$ fm. Both $R_{C}$ and $R_0$ are the parameters defining sizes of very different fields, the Coulomb and the nuclear potentials. In DWUCK programm $R_{C}$ is a rather formal parameter, which is a distance where the spherical function (10) inside the nucleus becomes equal the asymptotic part, while $R_0$ is the half-potential radius. Taking into account that protons, the source of the Coulomb field, are bound in a dense pack due to the nuclear force, one can suggest that the values $R_{C}$ and $R_0$ should increase with $Z$ in a consistent way. So one can suppose that $R_0\approx R_{exp}+d'$ where $d'\le d$.

The W-S potential satisfying the EPN-condition provides reasonable relation between $R_C$ and $R_0$ for the nuclei with $9\le Z\le 16$. Rms radii $<r_p^2>^{1/2}$ are in a good agreement with $R_p$ for the nuclei with $5\le Z\le$20. For the light nuclei with $Z\le 8$ the radii $R_C$ are too large, so that either no solution (in case of $\rm^{12}C$ and $\rm^{14}N$) or the value $<r_p^{2}>^{1/2}$ fm is less than the radius of the nucleus ($\rm^{16}O$). For heavier nuclei $Z\ge 17$ radii $R_0$ become unreasonably large in comparison with $R_{exp}$ and the difference $<r_p^2>^{1/2}-R_p$ starts increasing with $Z$.

When the criterion in finding proper parameters of the W-S Potential is the equation $<r_p^2>^{1/2}=R_p$, the fitted values $R^{DWUCK}_{C}$ are in an agreement with $R_C$ for the same nuclei $9\le Z\le 16$ and for the others $Z\le 8$ and $Z$> 16 there is disagreement with $R_C$. For light nuclei $R^{DWUCK}_{C}<R_C$ and for heavier nuclei $R^{DWUCK}_{C}>R_C$. The radius $R_{0}$ is comparable with $R_C$ but still the rate of its increasing with $Z$ is higher than it is supposed to be according to the rate of increasing of the experimental radii.

The values $R_C$ obtained by (6) may be too large for the light nuclei, because the approach of charge sphere is not good for the case. In this connection the values obtained by the criterion $<r_p^2>^{1/2}=R_p$ may be more appropriate. Using the W-S potential for the nuclei with $Z\ge 17$ in both cases of $R_C$ criterion and $R_p$ criterion  do not provide proper relation between the Coulomb radii, nuclear potential radii, and the rms radii of the last proton with the experimental radii.

The standard potential radius, which is taken equal to the standard Coulomb radius $R_C=R_0=1.25(A-1)^{1/3}$, is not good either, although rms radii are in some agreement with $R_p$ for all the nuclei. But the rate of increasing of $R_C$ and $R_0$ is not consistent with the rate of experimental radii increasing. For the nuclei with $10< Z\le 24$ the experimental radius $1.005(A-1)^{1/3}\le R_{exp}\le  1.077(A-1)^{1/3}$ and for the other nuclei with $Z\ge 24$ the experimental radii are  well described by the function $R^{A,Z}=1.005A^{1/3}$ ( Eq. (7) rewritten for $A=4N_\alpha$). It means that the standard potential radius $1.25(A-1)^{1/3}$ increases with $Z$ with a considerably higher rate than the radius of the nucleus, which is in a clear discrepancy with the short range nuclear force.  Besides, at the standard parameters the neutron and proton potential wells are allowed to be different. In case of the nucleus $^{30}$P the last neutron potential depth is less than the last proton one, $V_{0n}=57.14$ MeV and $V_{0p}=57.48$ ÌýÂ, which does not look right. The difference does not seem significant in point of view of the mean field theory where neutrons and protons are distributed independently and their centers of mass do not coincide. But it is not right in the representation of the $\alpha$-cluster model, where a nucleus consists of $np$-pairs joined in $\alpha$-clusters and the position of the last neutron and the last proton belonging to one pair is determined by one potential.

The question about consistency of the W-S potential used as a single particle potential with experimental radii in the point of view of charge distribution resulting from adding all protons distributions in the nucleus was discussed in a number of articles, for example \cite{2,7,8,9}. There was shown that a theoretical charge distribution calculated with using the sum of squared single particle wave functions produced by the W-S potential with a radius close to the standard one is in agreement with experimental charge density distribution for the nuclei with large deviation of $A$. But the sizes of internal shells are not known, so the fitting experimental radius can be done by varying sizes of internal shells. Besides, as it is already shown in section II experimental radius alone can not be a sensitive criterion for checking validity of the last proton rms radii calculated at the parameters. In spite of the fact that the proton rms radii for the nuclei with $Z>26$ considerably deviate from the $R_p$ the nuclear radii $R^{(A,Z)}$ (18) with using $<r_p^2>^{1/2}$ (recalculated for the cms of the residual nucleus with mass number $A-1$ (12)) instead of $R_p$ are consistent with $R_{exp}$, see FIG.\ref{fig1}. That is because the relative weight of the last proton decreases with $Z$. This is another evidence that the experimental nuclear radius alone can not be used as a sensitive test for validity of the last proton potential parameters without an additional criterion like $R_p$.

There is another remark about lack of consistency of the W-S potential with the nuclear density distribution. In the self-consistent calculations it is shown \cite{34} that the single particle potential should have no symmetry in the surface thickness to be consistent with the nuclear density. The internal part is to be considerably larger $t_{0.5}-t_{0.9}>t_{0.1}-t_{0.5}$ where $t_{0.5}$ means the half-potential radius.

If for the light nuclei with $Z\le 9$ the W-S potential is good, one has an opportunity to merge two approaches, the alpha cluster model and the wave function representation of a nuclear bound state, in case of heavier nuclei. In the framework of the $\alpha$-cluster model a nucleus is considered to consist of a core (an $\alpha$- cluster liquid drop with excess $nn$-pairs, which increases with $A$) and a molecule on its surface \cite{27,28}. According to this model the great majority of even $Z$ stable nuclei have a $\rm ^{12}C$ molecule on the surface of the core and odd $Z$ nuclei have on the surface a molecule $\rm^{15}N$. The last nucleon is supposed to be in the mean field of the molecule due to the short links. So, with increasing $A$ the core increases, but the last nucleon potential stays unchanged and the Coulomb potential for the last proton becomes larger. The parameters of the molecule potential can be found under the EPN condition for the nucleus corresponding to the molecule. The center of mass of the molecule is shifted from the center of mass of the whole nucleus by some distance $\Delta$ which increases with $A$. It will bring some other values of ANC and SF for heavier nuclei, because the wave function of the last nucleon will be restricted in the aria determined by the position of the molecule. Such representation will remove the mentioned above discrepancy and will provide a proper value $R_0=R^{ml}_0+\Delta$ where $R^{ml}_0$-the potential radius of the last nucleon in the molecule. Besides, the function will help in solving the long standing problem of selecting optical potential parameters used for describing of input and output elastic channels of the reaction, because the internal part of the amplitude of the reaction will be naturally cut off at small radii, which will make differences of the values of the parameters of different optical potential sets less important for the calculated cross sections. Technically the potential can be used in numerical solutions of the Shr\"{o}dinger equation. The parameter $\Delta$ can be estimated in the framework of the $\alpha$-cluster model or it can be found in the well-depth procedure at the other parameters fixed to adjust the experimental value of the single particle binding energy. A simple phenomenological proof of this representation can be found in the values of the experimental binding energy of the last neutron in the symmetrical nuclei heavier than the nucleus $^{12}$C. They group around two values $\varepsilon_n=$15.5  MeV for even $Z$ and $\varepsilon_n=$10.5 MeV for odd $Z$ and differences are obviously defined by spin-orbit correlation (see Table II).  So one can suggest that the last neutron has two kinds of links in dependence on whether there is a single $np$-pair or not. The energy of neutron separation in case of odd($Z$)-odd ($A$) stable nuclei (according to the model these nuclei consist of a core made of $\alpha$-clusters, $nn$-pairs placed in the core, and the molecule of $^{15}$N on the surface of the core) is also within few MeV around the value of the neutron separation energy in the nucleus $^{15}$N $\varepsilon_n=$10.5 MeV. For example, for the nuclei $^{19}$F, $^{27}$Al, $^{35}$Cl, $^{89}$Y, $^{141}$Pr and $^{209}$Bi $\varepsilon_n$=10, 13, 13, 12 and 9 MeV, which also says about similar conditions of the bond state of the last neutron determined by the links with the single $np$-pair and three $\alpha$-clusters in its close vicinity.

\section{CONCLUSION}
In the framework of alpha-cluster model the Coulomb radii $R_C$, RLPP $R_p$ were found for symmetrical nuclei with $5\le Z\le 45$. Relation between the Coulomb radius and nuclear radii is as follows $R_C\approx R_{exp}+d$ where $d\approx 1$ fm. Similar relation is expected for nuclear radius $R_0\approx R_{exp}+d'$ where $d'$ may not be equal to $d$. For the values of $R^{DWUCK}_C=R_C$ the W-S potential parameters satisfying the EPN condition in solving the Shr\"{o}dinger equation have been found for the nuclei with $9\le Z\le 16$. The rms radii for the last proton $<r^2_p>^{1/2}$ in case of the nuclei are in an good agreement with $R_p$. Taking into account that the obtained values $R_C$ for the light nuclei (i.e. $5\le Z\le 8$) can be too large, the parameter $R^{DWUCK}_C$ is used as a fitting parameter to fulfill the demand $<r_p^2>^{1/2}=R_p$ to find the parameters of the W-S potential. In case of pure peripheral reactions with one nucleon transfer reactions like $A(d,t)B$ or $B(^3{\rm He},d)A$ the error of spectroscopic factor obtained by means of the DWBA analysis with standard potential parameters is estimated. It is shown that the error that comes due to not proper parameters for some nuclei can be as large as $48\%$.
It has been shown that the standard potential increases with $Z$ at the rate much higher than it is supposed to be according to the rate of increasing of nuclear size, though for the nuclei $15\le Z \le 22$ they provide better agreement between rms radius $<r^2_p>^{1/2}$ and $R_p$ than the EPN parameters. This is provided by unreasonably big $R_0$ and the prevalence of the proton potential over neutron one, which does not look right.

Analysis of the values of $R_C$, $R_p$, the calculated at EPN-condition $R_0$ together with $R_{exp}$ shows that W-S potential is not proper for the task for the nuclei with $Z >$ 16.

A comparative analysis of the last neutron binding energies (see Table II) shows that the bound condition is defined by the short range interactions with the nearest presumedly three alpha-clusters and depends mostly on whether it is in the last $np$-pair (odd $Z$) or it belongs to the last alpha cluster (even $Z$). The single particle nuclear potential is to be of a shorter radius $R_0^m$ which is the potential radius of the nucleon within the $\rm ^{14}N$ or $\rm^{16}O$ molecule. Then it should be placed at some distance $\Delta $ from the center of mass of the whole nucleus. The distance $\Delta $ is defined by the radius of the core. In that case for heavier nuclei it could provide a proper relation between $R_C$, $R_p$, $R_{exp}$ and $R_0=R_0^m+\Delta $.

\end{document}